# Conductive Atomic Force Microscopy of Chemically Synthesized Graphene Oxide and Interlayer Conduction

Yoshio Kanamori,[1] Seiji Obata,[1] and Koichiro Saiki*[1,2]
[1]*Department of Chemistry, School of Science,* [2]*Department of Complexity Science and Engineering, Graduate School of Frontier Sciences, The University of Tokyo,*
*5-1-5 Kashiwanoha, Kashiwa, Chiba 277-8561*

Graphene oxide, one of chemically modified graphenes, has been attracting wide attention because of promising adaptability to a wide variety of applications. However, the property of graphene oxide itself has not been known well. Using a conductive cantilever, we observed a current image of graphene oxide nanosheets with various thicknesses. Current –voltage characteristics were found to reflect the local conductivity normal to the nanosheets. Under high electric fields, the conduction was well described in terms of Poole-Frenkel emission mechanism. The fitting of *I-V* curves to the Poole-Frenkel model provides information on dielectric property and the relative permittivity of graphene oxide was evaluated to be 4.8 ± 0.8.

Recently, graphene has attracted great attention due to its unique electronic, mechanical, and thermal properties. Several synthesis methods of graphene have been reported, including mechanical exfoliation of bulk graphite[1], chemical vapor deposition (CVD)[2,3], and reduction of graphene oxide (GO)[4]. Among these methods, the route via GO is considered to be appropriate for large volume production, in which graphite powder is oxidized and GO sheets are dispersed in the solution. Apart from graphene, GO and its derivatives are expected to be applied to transparent electrodes in electroluminescence or photovoltaic devices[5], memory FETs[6], and catalysts[7].

The preparation method of GO and the properties of reduced GO have been studied intensively. Plenty of literatures have been published these few years.[4] With regard to the structure or chemical composition of GO, there has been a consensus that several types of functional groups such as epoxide, hydroxyl, carbonyl, carboxyl, etc. exist in the plane of GO and the effective thickness is larger than that of graphene. In contrast to the above works, only a limited number of reports are available concerning the electrical or optical property of GO. The band gap of GO[8] and the refractive index of GO[9] were reported previously.

In the present work, we observed morphology of GO by use of atomic force microscopy with a conductive cantilever. Topographic and current images were obtained simultaneously. The similarity between both images indicates that current does not diffuse laterally beyond the spatial resolution of images (0.05 μm) but flows normally to the GO nanosheets. Thus dependence of the conductivity on the electric field and the GO thickness provides information on electric transport perpendicular to the graphene sheets of GO. Interlayer conduction of GO nanosheets could be explained in terms of Poole-Frenkel model in high electric fields.

Graphene oxide (GO) was prepared from natural graphite powder (SEC carbon) by the modified Hummer's method[10,11]. Briefly, graphite powder, $NaNO_3$, and $H_2SO_4$ were placed in a flask and the mixture was stirred while being cooled in an ice/water bath. $KMnO_4$ were then added gradually for about 1 h. Cooling was completed in 2 h, and the mixture was allowed to stand for 5 days at 20 °C with gentle stirring. The oxidized product was purified by rinsing with $H_2SO_4$, and $H_2O_2$ solutions repeatedly. In our experiment, GO flakes were dispersed in methanol after solvent substitution. The solution was centrifuged and the final product was supernatant fluid containing GO sheets with various thicknesses (monolayer to a few layers). A highly doped Si substrate was then dipped in the dispersion liquid and lifted. Mild centrifugation during the preparation process or repeated dipping yielded thicker GO sheets on the Si substrate[12]. The final products were characterized by XPS. The atomic ratio of O/C was estimated to be 0.4~0.5, which is in good agreement with the previous report[4].

Topographic and current images of GO sheets were measured with a scanning probe microscope (JEOL JSPM-5200). A schematic diagram of the measuring system is shown in Figure 1(a). A Si substrate with GO was set on the mica plate attached to the piezo stage. By applying a bias voltage to the Si substrate, topographic and electric current images were obtained simultaneously. The current was measured by a pre-amplifier to the order of picoamperes. The images were obtained under high vacuum conditions ($<10^{-3}$ Pa) to prevent the contamination of the probe. After measuring the topographic and current images, the cantilever was kept at a certain point on the GO sheet to measure the current-voltage characteristics. The measurement was carried out by a source meter (Keithley model 6487) under ambient conditions. The contact force during *I-V* measurement was kept constant at 10 nN. The current gradually decreased during measurement, possibly because of the probe contamination by application of high voltages. Therefore, the cantilever was replaced by a new one before each measurement.

First, the flatness of GO sheet was examined. In order to compare GO with $SiO_2$, GO was deposited on a $SiO_2$ (300 nm)/Si substrates. A typical AFM image of GO, which was measured in non-contact mode, is shown in Figure 1(b). A height distribution which was measured both for GO and $SiO_2$ within the 300 nm squares (surrounded by white dashed lines in Figure 1 (b)) is shown in Figure 1(c). Both curves are almost similar to each other, indicating that the GO surface is as flat as that of $SiO_2$. The surface roughness of GO on $SiO_2$ is 0.2 nm, which is almost the same as the graphene on $SiO_2$[13].

Contact AFM topographic images and current images of GO layers are depicted in Figure 2. A topographic image of a single and double GO layers are shown in Figure 2(a) together with the height profiles taken along the dashed lines (i) and (ii). The line (i) crosses a single GO step, the height of which is around 0.8 nm. The line (ii), on the other hand, crosses over the folds of GO sheet or the wrinkles on the GO surface, the height of which reaches a few nanometers[14]. The current image of the same area, which was measured at a bias voltage of 5 V, is shown in Figure 2(b). Brighter areas correspond to higher electric current, in which the boundary between a single and double layer is clearly seen. It is noted that the spatial resolution in the current image is as high as that in the topographic image. Figure 2(c) shows a topographic image of GO aggregates. Thicker GO sheets ranging from 5 nm to 35 nm are observed. The corresponding current image measured at a bias voltage of 10V is shown in Figure 2 (d). The contrast in the thicker region (bottom left) seems enhanced, while that in the thin region (bottom right) disappears. Thus, the contrast in the current image can be controlled by the applied voltage, taking account of the GO thickness. The results of Figure 2(b) and (d) show that the current images have resolution as high as the topographic images. This means that the current does not diffuse parallel to the GO sheet beyond the lateral resolution (less than 0.05μm), which is in close agreement with the fact that the as grown GO is insulating for the electric field parallel to the sheet[8].

After getting the topographic and current images, the cantilever was set at various positions on the flat GO sheets with different thicknesses, where local current-voltage (*I-V*) characteristics were measured. *I-V* characteristics of GO layers thinner than 8 nm showed fluctuations probably because the current is very sensitive to contact force.[15] The *I-V* characteristics of the GO layers thicker than 8 nm, however, were measured reproducibly and reversibly. Therefore we focus on the GO layers thicker than 8 nm in the following discussion.

The inset of Figure 3 shows *I-V* characteristics, which indicates a strong nonlinear behavior with applied voltage. The conduction mechanism with such a non-ohmic feature is often ascribed to Schottky emission mechanism or Poole-Frenkel emission mechanism, which originates from lowering of potential barrier for carriers out of a defect center or a trap under an applied electric field. It has been reported that *I –V* curves in thin dielectrics can be fitted to a Schottky emission mechanism at low electric fields and a Poole–Frenkel conduction mechanism at high electric fields[16]. The current originating from Poole-Frenkel emission is described by Eq 1,

$$I \propto F \exp\left(\frac{\sqrt{q^3 F / 4\pi\varepsilon}}{kT}\right), \qquad (1)$$

where $F$ is an electric field applied normal to the GO sheet, $\varepsilon$ is permittivity of GO, $q$ is a unit charge, and $T$ is temperature. *I-V* characteristics are then plotted in semilog ($I/F$) vs. ($F^{1/2}$) for the GO layers with various thicknesses. At high electric fields, most of points are on the straight lines of the same slope, independently of the GO thickness from 8 nm to 50 nm. The good linear fit in semilog ($I/F$) vs. ($F^{1/2}$) plot indicates that the electric conduction normal to the GO layers at high electric fields is ascribed to Poole-Frenkel emission mechanism, although the deviation from the linear fit at low electric fields may originate from the contribution of Schottky emission mechanism. As the GO thickness increases, the ($I/F$) vs. ($F^{1/2}$) curve deviates downwards from the universal curve. The origin of the deviation is not definite at the present stage, while the decrease in effective electric field might be one of the reasons. Since the diameter of probing cantilever is around 50 nm[17], the electric field extends beyond the electrode for the GO film thicker than a few tens of nanometer. For the thinner film, on the other hand, the electric field is confined just below the probing cantilever like a parallel plate capacitor.

The Poole-Frenkel type conduction has been reported in various insulators; $SiO$[18], $ZrO_2$[16], $HfO_2$[19], and other materials. In these materials the donor-like electron trap sites are distributed continuously in the materials. In the case of GO, however, electrons are trapped in each GO layer and they transit above the potential barrier between layers. In order to evaluate the trapping barrier height, temperature dependent *I-V* measurement is required, which could not be achieved in the present work due to limitation of instrumental capabilities.

Since the experimental *I-V* characteristics at high electric fields is well fitted by Poole-Frenkel model, the relative permittivity of GO can be estimated using Eq 1. It is worth noting that the relative permittivity in Eq 1 corresponds to that at the high frequency limit, or the square of refraction coefficient. Fitting ($I/F$) – ($F^{1/2}$) plots by eq. (1), the relative permittivity is calculated to be $4.8 \pm 0.8$ (assuming $T$ = 293 K). This value is almost consistent with the calculated permittivity ($\varepsilon_r$ = 3.23~4.76), based on the index of refraction and absorption of GO reported in Ref. 9.

The results mentioned above were obtained applying a positive bias voltage to the GO nanosheets. The dependence of current on the electric field and the GO thickness under negative bias is almost the same with that under positive bias, although the current intensity was slightly lower under the negative bias.

In summary current images of GO nanosheets with various thicknesses were observed for the first time by a conductive cantilever of AFM instrument. The current image is useful to monitor the GO thickness as well as wrinkles and folds in the nanosheet with high contrast. The current-voltage characteristics normal to the GO layers is well described in terms of Poole-Frenkel emission mechanism and the permittivity of GO can be evaluated from the curve fitting.

This work was supported by a Grant-in-Aid for Scientific research (No. 21360005) from the MEXT of Japan. One of the authors (Y.K.) acknowledge Global COE Program “Chemistry Innovation through Cooperation of Science and Engineering”, MEXT, Japan, for their financial support.

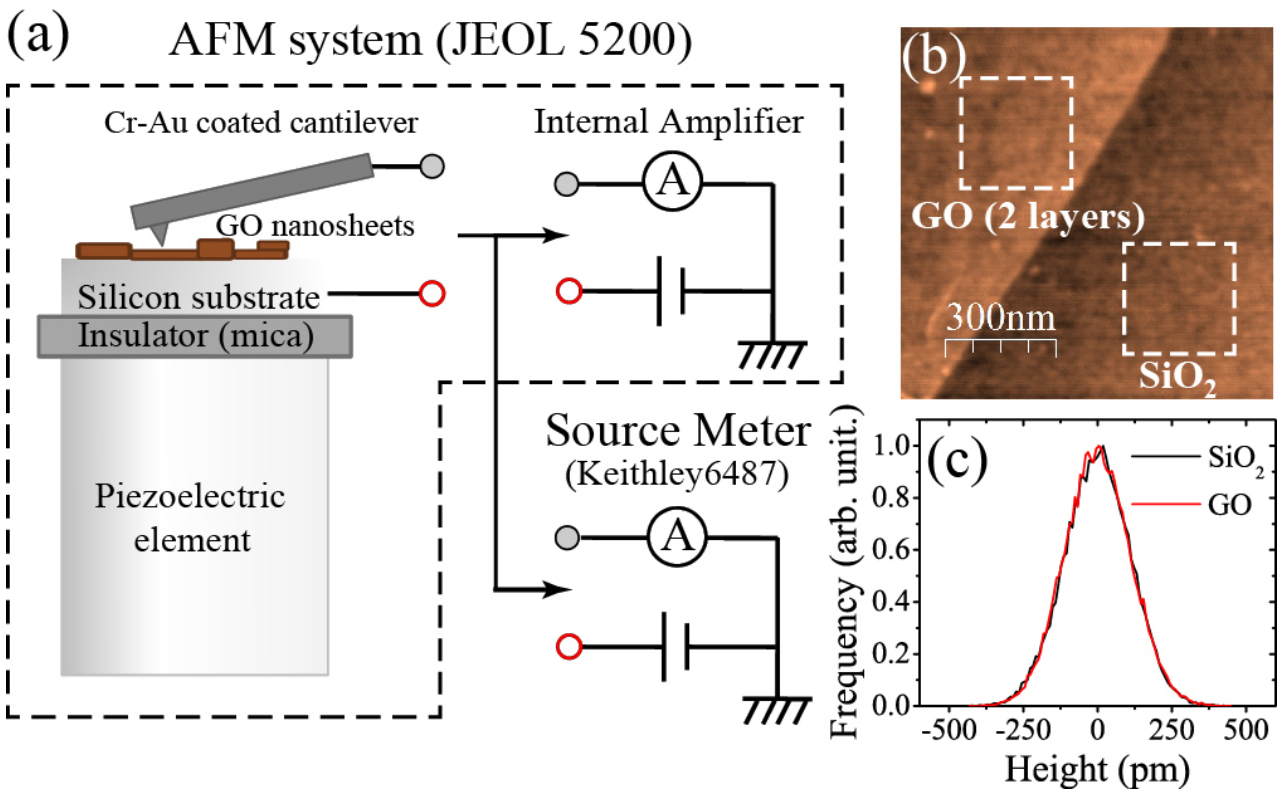

**Figure 1.** (a) A schematic diagram of the measurement system. The cantilever is connected to an internal amplifier for obtaining the current image and to a source meter for measuring *I-V* characteristics. (b) A non-contact AFM image (1μm × 1μm) of 2 layers GO. (c) Height distributions of GO and $SiO_2$ surfaces, which are measured in squares surrounded by white dashed lines in (b) (300nm × 300nm).

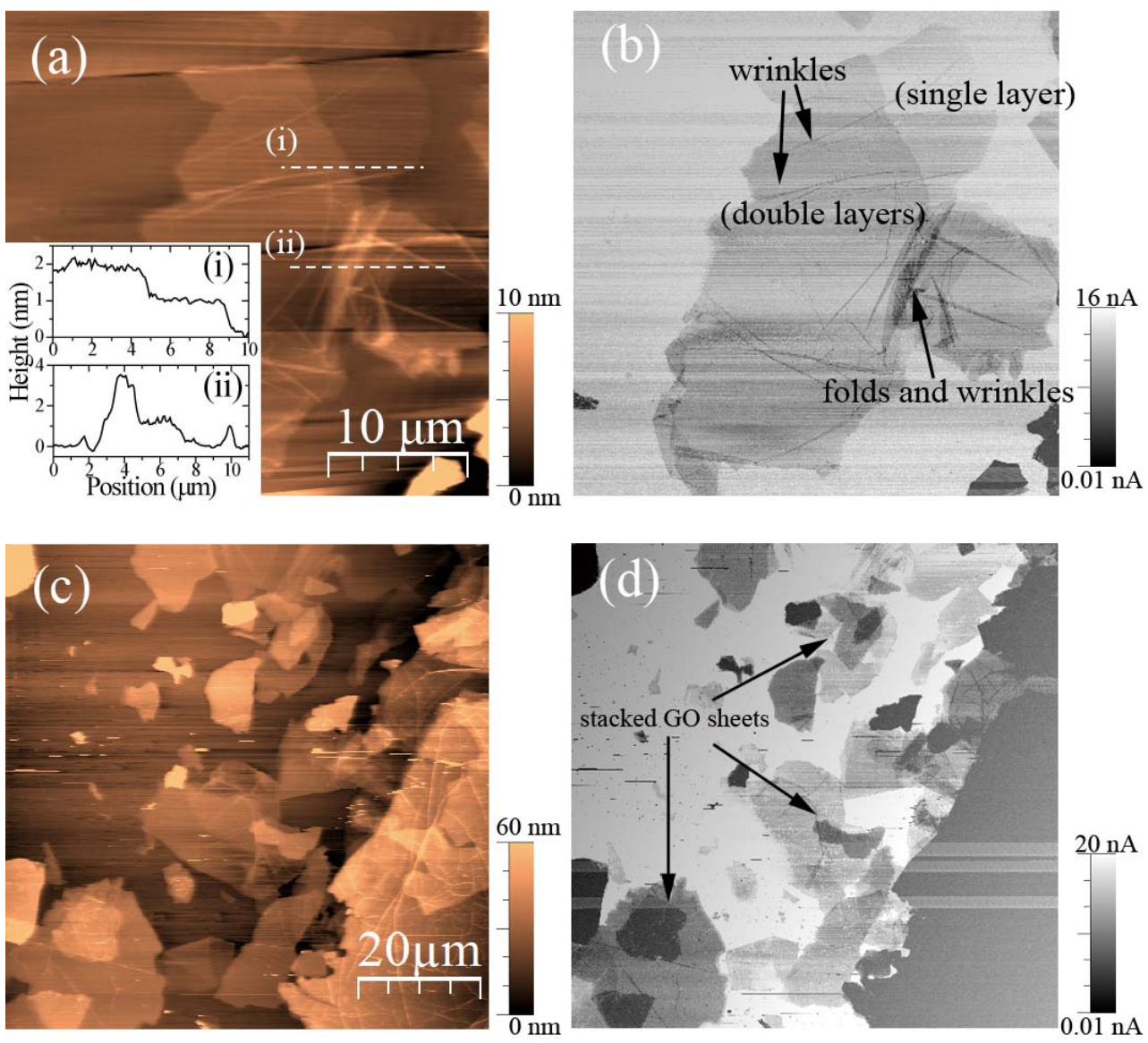

**Figure 2.** (a) A contact AFM image (35 μm x 35 μm) and (b) a current image of a few layers GO. Inset in (a) shows the height profiles taken along the white dashed lines. (c) A contact AFM image (80 μm x 80 μm) and (d) a current image of GO aggregates. The current image was taken with the sample-to-tip bias of +5 V (b) and +10 V (d).

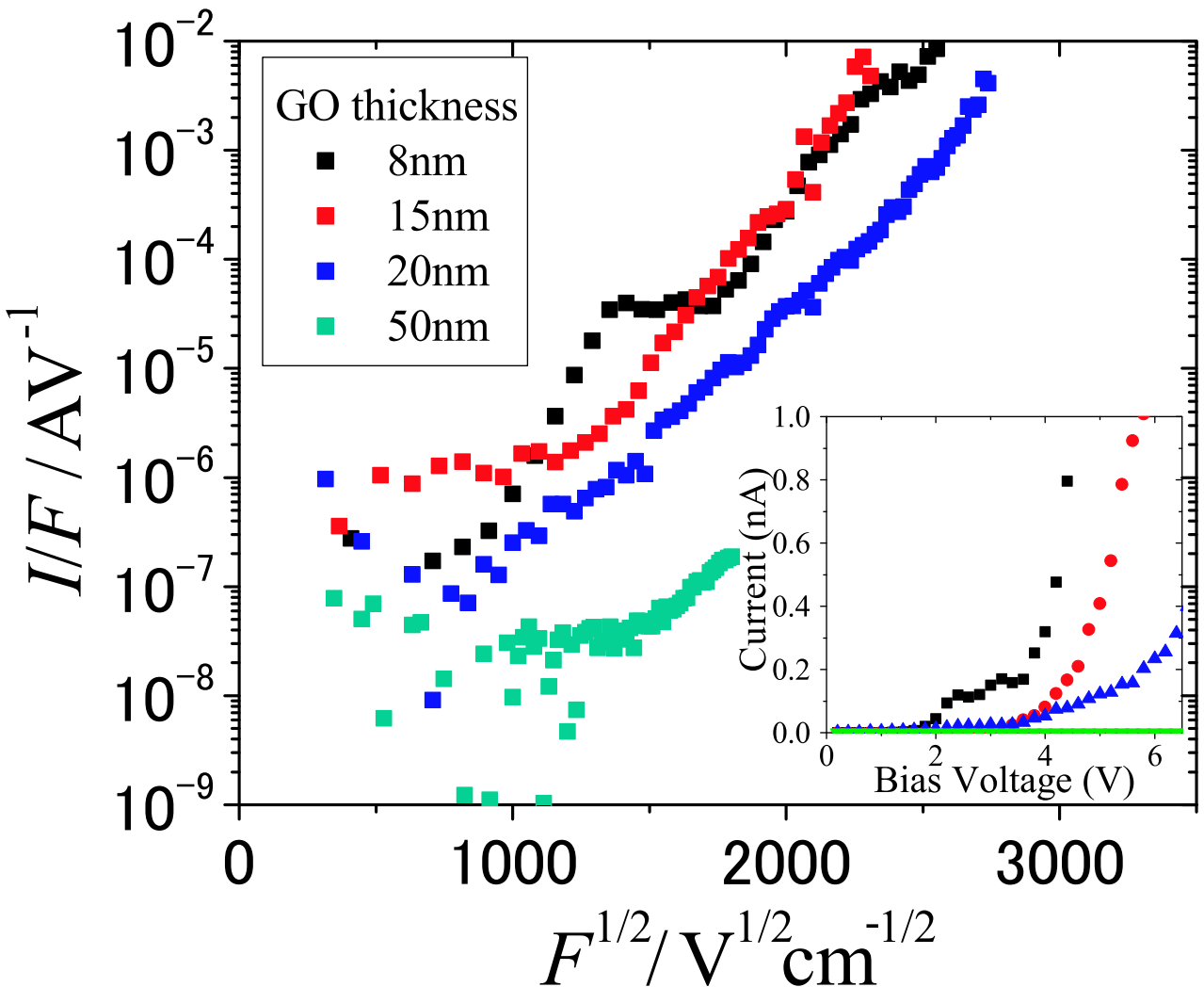

**Figure 3.** *I-V* characteristics of GO sheets with various thicknesses. The inset shows *I-V* curves in linear scale.